\documentstyle[12pt,aaspp4]{article} 
\lefthead{Outram, Carswell & Theuns} 
\righthead{Non-Voigt Lyman-alpha absorption line profiles}

\begin{document}
\title{Non-Voigt Lyman-alpha absorption line profiles}
\author{P. J. Outram\altaffilmark{1} and R. F. Carswell}
\affil{Institute of Astronomy, Madingley Road, Cambridge CB3 0HA,
U.K.}  
\and  
\author{T. Theuns} 
\affil{Max-Planck-Institut f\"{u}r Astrophysik, 
Karl-Schwarzschild-Str. 1, D-85740 Garching bei M\"{u}nchen, Germany} 
\altaffiltext{1}{now at: Department of Physics, University of Durham, 
Duham DH1 3LE, U.K.}

\begin{abstract} 
Recent numerical simulations have lead to a paradigm shift in our
understanding of the intergalactic medium, and the loss of a physical
justification for Voigt profile fitting of the Lyman-alpha
forest. Many individual lines seen in simulated spectra have
significant departures from the Voigt profile, yet could be well
fitted by a blend of two or more such lines. We discuss the expected
effect on the line profiles due to ongoing gravitational structure
formation and Hubble expansion. We develop a method to detect departures
from Voigt profiles of the absorption lines in a statistical way and
apply this method to simulated Ly$\alpha$\  forest spectra, confirming
that the profiles seen do statistically differ from Voigt profiles.
\end{abstract}

\section{Introduction}
Recent hydrodynamic simulations of large-scale structure formation in
cold dark matter (CDM) dominated cosmologies (e.g. Cen et al. 
\markcite{cen94} 1994,
Zhang, Anninos \& Norman \markcite{zha95} 1995, 
Hernquist et al. \markcite{her96} 1996, Theuns et
al. \markcite{the98} 1998) have been able to naturally 
reproduce the Lyman-alpha
(Ly$\alpha$) forest. These simulations have been remarkably accurate
in reproducing the column density, and Doppler width distribution, as
well as evolutionary properties seen in the Ly$\alpha$ forest of real
QSO spectra (e.g. Kirkman \& Tytler \markcite{kir97} 1997, 
Lu et al. \markcite{lu96} 1996).

Early attempts to model the  Ly$\alpha$ absorption lines considered
pressure-confined clouds (Sargent et al. \markcite{sar80} 1980, 
Ostriker \& Ikeuchi \markcite{ost83}
1983) inside a hot intergalactic medium (IGM). Other ideas used dark
matter mini-halos to gravitationally confine the Ly$\alpha$ clouds
(Rees \markcite{Ree86} 1986). 
In these models a Voigt profile naturally arises from a
Maxwellian thermal distribution, or Gaussian turbulent motion of the
gas. This lead to the fitting of Voigt profiles to the data; a method
now generally used to analyse the complex blended features seen in the
high resolution Keck spectra (e.g. Kirkman \& Tytler \markcite{kir97}
1997).

The numerical simulations have brought about a paradigm shift in our
understanding of the IGM. No longer is the Ly$\alpha$ forest viewed as
many isolated objects within a hot tenuous IGM, but rather it is seen
as the density fluctuations of the IGM itself caused by gravitational
collapse into a network of sheets and filamentary structures
(sometimes described as a fluctuating Gunn-Peterson effect, after 
Gunn \& Peterson \markcite{gun65}
1965). The observed spectrum is thus described by the
density, temperature and velocity of the gas at all points along the
line-of-sight, and the absorption lines that make up the forest are
caused by extended over-densities, rather than discrete clouds.

This change in our understanding of the IGM has called into question
the justification behind Voigt profile fitting. In recent years,
dramatic improvements in the resolution and signal-to-noise ratio
(S/N) of data has lead to complex blends of Voigt profiles in order to
fit the absorption features. Problems due to the somewhat arbitrary
number of components, and the non-uniqueness of the fitted solutions
have become apparent. Now the simulations have shown that the physical
interpretation of the solutions themselves are in doubt. Despite this
there is little direct evidence that the Ly$\alpha$ forest line
profiles are significantly different from Voigt profiles. Many
absorption lines appear to have remarkably Gaussian velocity
distributions, especially at lower redshift where blending is less of
a problem. On the other hand, non-Gaussian features could as well be
attributed to blends of two or more Voigt profiles as to an
intrinsically non-Voigt distribution.

Outram et al. \markcite{out99} (1999) proposed a method to detect 
departures from Voigt
profiles of the absorption lines in a statistical way. They applied it
to the Ly$\alpha$\  forest spectrum of GB1759+7539, but detected no
significant evidence of non-Voigt profiles. In this letter we develop
the method proposed by Outram et al.. In the next section we discuss
the signature of non-Voigt profiles in Ly$\alpha$\  forest spectra, and
present details of the method to detect it. Then we apply this method
to simulated forest spectra to show that the profiles seen in
hydrodynamic simulations are indeed statistically non-Voigt, before
finally discussing the implications of this result.

\section{Detecting Non-Voigt Profiles}
\begin{figure}
\figurenum{1}
\plotone{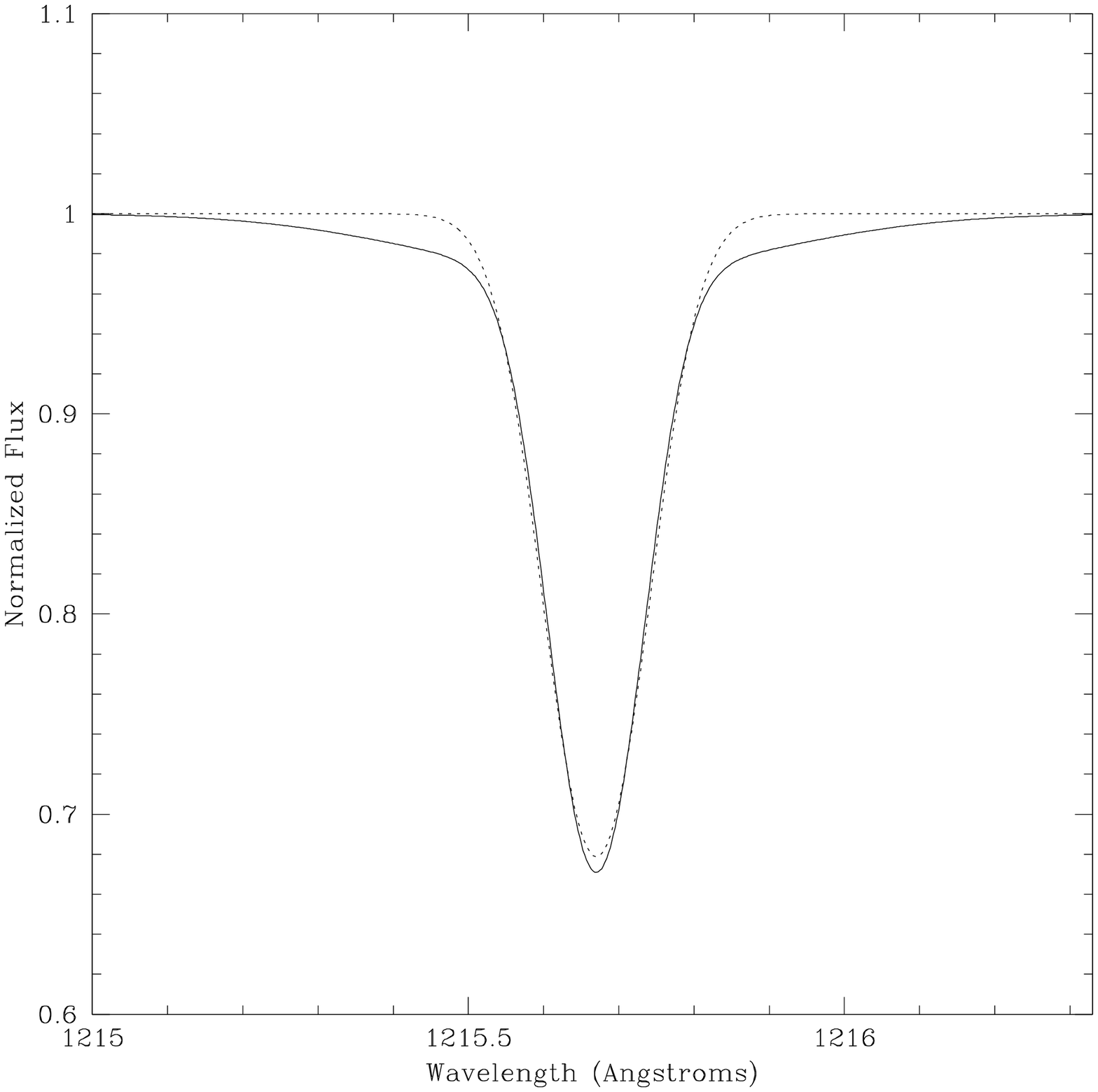}
\caption{Example non-Voigt absorption line profile with broad non-Maxwellian wings and a narrow core. The dotted line shows a Voigt profile fit to this feature.
}
\label{fig1}
\end{figure}

\placefigure{fig1}

If the Ly$\alpha$\ forest is now viewed as density fluctuations within
the IGM, then what does this imply about the physical nature of
Ly$\alpha$\  absorbers? Typical systems have column densities of
around $13.0 < $log $N$(H$\:${\small I})$ < 14.0$
(log$\:$cm$^{-2}$). Observations of quasar pairs have shown that these
systems can be very large; of the order of 500 kpc across (Dinshaw 
et al. \markcite{din94} 
1994). They are therefore highly ionized, and hence contain a
significant fraction of all the baryons at $z=3$, yet they are only
slightly overdense ($\rho/\bar{\rho}\sim 1-10$) compared to the mean
baryonic density. Although a large variety of structures are seen in
the simulations, they tend to have a flattened geometry, in the form
of ``pancakes'' or filamentary structures. They are unlikely to be
virialized objects, and are probably transient density fluctuations
undergoing gravitational collapse in one direction, whilst still
expanding with the Hubble flow in others (Haehnelt \markcite{hae96} 1996).

The individual absorption profiles depend on the orientation and exact
geometry of that system, and may well be asymmetric. In general
though, if Ly$\alpha$\  absorbers are in a state of collapse then the
bulk motion, and compressional heating of the gas should lead to broad
non-Maxwellian wings (Rauch \markcite{rau96} 1996,
Nath \markcite{nat97} 1997). Equally if the absorber
is extended in space, and undergoing Hubble expansion along the
line-of-sight then the line profile would also deviate from that
predicted by the simple model.

The expected shape of a typical absorption line from such objects is
shown in figure 1. The central core is that of a log $N$(H$\:${\small
I})$ = 13.0$, $b = 20$ km$\,$s$^{-1}$ absorption line, with broad
non-Maxwellian wings due to infalling gas. When a spectrum is fitted
with Voigt profiles, these wings may be fitted by a single broad, low
$N$(H$\:${\small I}) component, or perhaps by two low $N$(H$\:${\small
I}) components, either side of the main line, or simply not fitted at
all. Rauch \markcite{rau96} (1996) considered the first of these possibilities,
searching for broad lines fitted simultaneously in redshift space with
narrower ones. He looked at both real and simulated data for an
anti-correlation of Doppler parameters for profile pairs at small
separations, detected a positive signal in both data-sets and
concluded that there was evidence of a departure from Voigt profiles.

In order to investigate the line profiles further, we estimate the
departures from the Voigt profile in absorption lines using the following 
method. Firstly,
for the spectrum in question, Voigt profiles are fitted to all
absorption features. Since we are looking for departures from Voigt
profiles in stronger lines, the raw spectrum is divided by an
artificial spectrum made by inserting the Voigt profile fits on the
continuum. The obvious thing to do is to divide through by the fitted
Voigt profiles, only leaving any non-Voigt residuals in the
spectrum. However, this is complicated by the fact that many of these
residuals will have been fitted themselves, using a blend of low
column density lines with high or low Doppler width, and therefore the
non-Voigt signal could be removed as well. In an attempt to overcome
this, all the systems with log $N$(H$\:${\small I})$ < 12.5$ are left
in for the examples here, though the precise limit can be any
desired. Although this leaves in many randomly-distributed small
absorption features, together with the non-Voigt residuals of removed
systems, the final step is then to co-add many of the absorption
systems whose Voigt core had been divided out. Any signature of
non-Voigt profile should be reinforced with this stacking, whereas
randomly distributed small absorption lines should be averaged away if
enough lines are stacked.

\begin{figure}
\figurenum{2}
\plotone{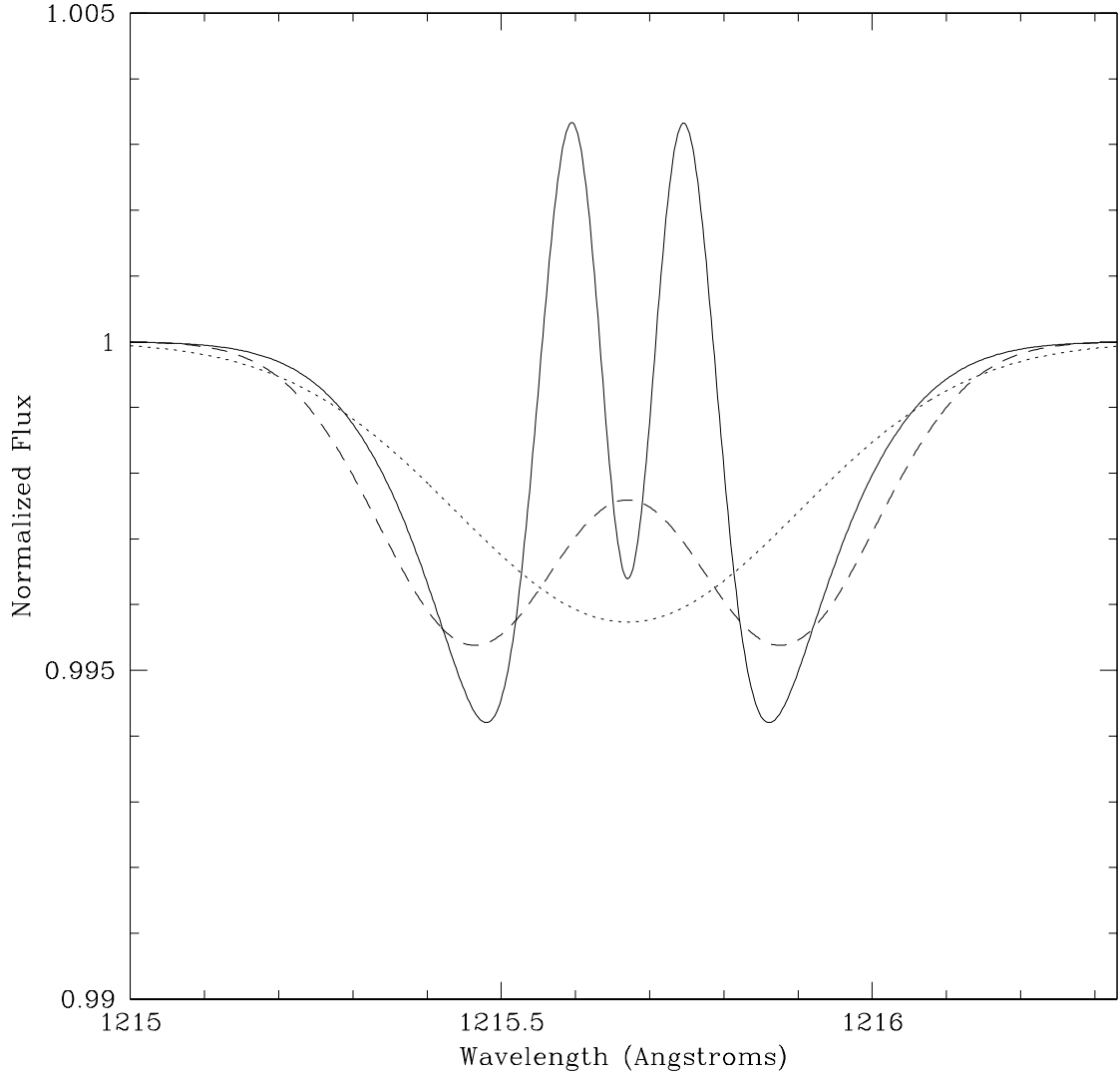}
\caption{Schematic residual spectrum. The solid line shows the expected residual if no Voigt profiles are fitted to the non-Voigt wings. The dotted line shows the residual if a single broad line were fitted, and the dashed line shows the residual if two narrower lines were fitted to the wings.}
\label{fig2}
\end{figure}

\placefigure{fig2}

The expected residuals using this method depend on how many components
are used to fit the non-Voigt wings. The general shapes can be seen in
figure 2. If the solid profile shown in figure 1 is fitted by a single
Voigt profile, which is then divided out, then the residual expected
is the solid curve in figure 2. As expected there are dips due to the
two wings, and inside these are two peaks due to the fact that the
Doppler width of the fitted Voigt profile is forced to be wider than
that of the core. If one or two extra lines are fitted to the wings
then the residuals would look like the dotted and dashed curves in
figure 2 respectively. When stacking hundreds of lines, a combination
of all three effects would be expected, and the ratio of each would
depend on the S/N of the data, and the fitting criteria used. The
randomly-distributed small lines would also have an effect; depressing
the entire spectrum by about 0.5\%. Within about 20 km$\,$s$^{-1}$ of
the absorption line centre (1215.60 - 1215.74\AA) the division to
remove the profile core becomes uncertain, especially for the higher
column density lines, where the residual signal is near zero. However,
since we are looking for a signal in the wings, as opposed to the
core, this region was ignored for the analysis.

\section{Simulated Ly$\alpha$\ Forest Profiles}

In order to test this idea further we used artificial spectra taken
from Theuns et al. \markcite{the98} (1998). The spectra were created 
using a simulation
code based on a combination of a heirarchical
particle-particle-particle-mesh (P3M) scheme for gravity and smoothed
particle hydrodynamics (SPH) for gas dynamics. The simulations assume
a standard adiabatic, scale-invariant CDM cosmology ($\Omega=1$,
$\Omega_{\Lambda}=0$, $H_0 = 50$ km$\,$s$^{-1}\,$Mpc$^{-1}$, $\sigma_8
= 0.7$, $\Omega_B = 0.05$). A box size of 5.5 Mpc was used, with
$64^3$ SPH, and an equal number of dark matter particles. Finally, the
assumed background radiation spectrum was half the amplitude of the
spectrum computed by Haardt \& Madau \markcite{haa96} (1996). For further 
details,
refer to Theuns et al. \markcite{the98} (1998) where this simulation is named
A-5-64. Theuns et al. tested the convergence of this simulation by
comparing it to a similar run with even higher numerical resolution
(A-2.5-64). They concluded that the A-5-64 run is very close to
convergence and that conclusions drawn from this simulation are
reliable.

We took the absorption spectra computed from 128 different
lines-of-sight through the box at redshift $z=3$. Each spectrum was
convolved with a Gaussian profile with full width at half maximum,
FWHM = 7 km$\,$s$^{-1}$, and Gaussian noise was added with standard
deviation $\sigma = 0.02$ (S/N=100 for pixels at the continuum) to
mimic spectra observed using the HIRES spectrograph on the Keck
Telescope. The spectra were wrapped around to enable fitting of
features that were otherwise close to the edge of the region.

Voigt profiles were fitted, using a $\chi^2$ minimization technique,
to the absorption features in order to determine the redshifts, column
densities and Doppler widths of the Ly$\alpha$\  absorption lines,
using an automated version of the software package VPFIT (Webb 
\markcite{web87} 1987;
Rauch et al. \markcite{rau92} 1992).

\begin{figure}
\figurenum{3}
\plotone{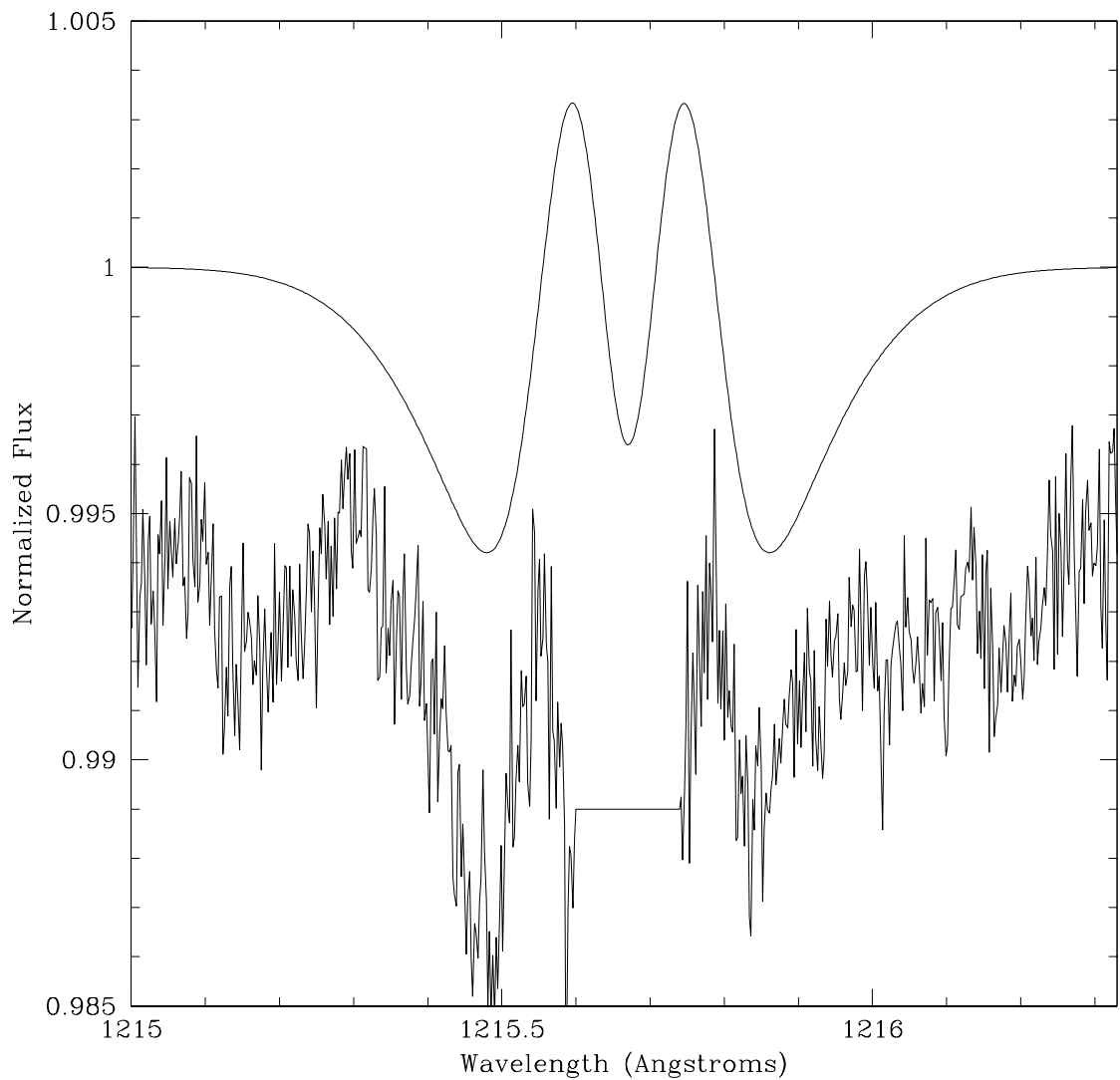}
\caption{Co-added residual spectrum obtained by stacking the residuals from 235 Ly$\alpha$\ lines, taken from simulated spectra at $z=3$, with $13.0 < $log $N$(H$\:${\small I})$ < 14.0$, and $15.0 < b < 60.0$, after first removing all fitted Voigt profiles with log $N$(H$\:${\small I})$ > 12.5$. The region from 1215.60 - 1215.74\AA\  is uncertain due to the division, and hence has been left out. The resulting S/N of the remaining spectrum is 680. The overlying curve shows a simulated non-Voigt residual.}
\label{fig3}
\end{figure}

\placefigure{fig3}

When comparing simulated data to observed spectra, one of the major
problems to be overcome is that of continuum fitting. The simulated
spectra show typically around 1-2\% zero-order absorption, which would
be removed by the continuum fitting procedure for real data. This is
usually done by using low-order polynomial fits to apparently
unabsorbed parts of the spectrum. The regions over which the continuum
is fitted are typically many times the size of a simulated
spectrum. In an attempt to overcome this problem, and treat the
simulated data in a similar manner to real data, VPFIT has been
developed to simultaneously fit a linear multiplicative factor to the
initial assumed continuum (unity for the simulations) during the
$\chi^2$ minimization procedure. The continuum was lowered by an
average of 1.6\% during the fitting of the artificial spectra.

The spectra were then divided through by the profiles of the stronger
fitted  lines (log $N$(H$\:${\small I})$ > 12.5$), leaving the
residuals due to weaker lines in. Care was taken to give zero weight
to those regions where the residual flux was below 20\%, and correct
the errors for the other regions accordingly.

The resulting residual spectra were co-added, weighted according to
variance, centred on the rest wavelength of the removed Voigt profile
absorption lines with parameters $13.0 <$ log $N$(H$\:${\small I})$ <
14.0$, and $15.0< b <60.0$ km$\,$s$^{-1}$. The result, with 235 lines
stacked, can be seen in figure 3.

If a line profile was indeed narrow in the centre, but with
broader-than-Maxwellian wings, the residual profile after fitting with
a Voigt profile, then dividing through by this fit would be as shown
by the smooth surve in figure 3 (modelled by a $b=20$ km$\,$s$^{-1}$,
log $N$(H$\:${\small I})$ = 13.0$ line with broader wings simulated by
a coincident $b=60$ km$\,$s$^{-1}$,  log $N$(H$\:${\small I})$ = 12.0$
line). This pattern seems to follow that seen in the residual spectrum
remarkably well. The dips at 1215.50 \& 1215.85\AA, and the peaks at
1215.55, and 1215.85\AA\ are similar to those predicted by the model
curve. The entire spectrum has been depressed by about 0.5\% due to
the small (log $N$(H$\:${\small I})$ < 12.5$) features not removed,
and these could also explain the small irregularities away from the
line centre. These irregularities would be expected to diminish as
more lines are stacked. Were the profiles of the absorption features
intrinsically random blends of Voigt profiles, then no such residual
features would be expected in the co-added spectrum. This is therefore
clear evidence of intrinsic departures from the Voigt
profile in the simulated spectra.

\section{Discussion}

The next step is to apply this method to real data in order to
determine whether the non-Voigt absorption profiles are a
simulated phenomenon, or a real physical property of the
absorbers. The stacked residual spectrum in figure 3 was created using
235 separate lines at S/N=100. This is of the same order as the number
of observed lines in a single high quality Keck spectrum, and so a
result should be easily achievable. Care would need to be taken to
remove regions where heavy-element absorption is detected, and the
continuum should be treated in a similar manner in order to produce a
fair comparison. The latter would mean fitting the spectrum in very
small chunks ($\sim 500$ km$\,$s$^{-1}$), simultaneously introducing a
local fit to the continuum in a similar manner to that described
above. The S/N of the spectrum is also important as it could change
the way that the residuals are fitted and hence the resulting residual
profiles, as discussed above.
 
 It has been noted that the absorption line profiles of individual
 systems in simulated spectra often appear to have broad wings or
 asymmetries, signifying a non-Voigt profile (e.g. Dav\'e et
 al. \markcite{dav97} 1997). Rauch \markcite{rau96}
(1996) showed that pairs of lines with small
 separations have anti-correlated Doppler-widths, suggesting that the
 Voigt profile decompositions are not actual blends. We have
 introduced a method to test whether or not the line profiles in the
 Ly$\alpha$\  forest are intrinsically non-Voigt. A similar technique
 was applied to GB1759+7539 (Outram et al. \markcite{out99}
 1999) with no sign of any
 signal. However, a much more extended sample is needed, with care
 paid to continuum uncertainties before the results can be
 compared. If wings are found in the real data, it will provide a
 powerful confirmation of the SPH models. If they are not, then they
 provide a crucial test and we need the modellers to think again.

\acknowledgements{The data analysis was performed on the
Starlink-supported computer network at the Institute of Astronomy. PJO
acknowledges support from PPARC and Queens' College.}

\newpage

\end{document}